\documentclass[graphicx, 12pt]{article}
\usepackage{indentfirst}

\usepackage{graphicx}
\usepackage{caption}
\usepackage{subfigure}

\begin{document}

\small
\hoffset=-1truecm
\voffset=-2truecm
\title{\bf The neutrino pair annihilation
($\nu\overline{\nu}\longrightarrow e^{-}e^{+}$)around a massive
source with an $f(R)$ global monopole}

\author{Yuxuan Shi, Hongbo Cheng\footnote
{E-mail address: hbcheng@ecust.edu.cn}\\
Department of Physics,\\ East China University of Science and
Technology,\\ Shanghai 200237, China\\
The Shanghai Key Laboratory of Astrophysics,\\Shanghai 200234,
China}

\date{}
\maketitle

\begin{abstract}
In this work we investigate the neutrino pair annihilation around
a gravitational object involving an $f(R)$ global monopole. We
derive and calculate the ratio $\frac{\dot{Q}}{\dot{Q_{Newt}}}$
meaning that the energy deposition per unit time is over that in
the Newtonian case. It is found that the more deviation from
general relativity leads more energy to set free from the
annihilation with greater ratio value. It should also be pointed
out that the existence of global monopole makes a sharp increase
in the ratio $\frac{\dot{Q}}{\dot{Q_{Newt}}}$, causing heavier
gamma-ray burst. We also discuss the derivative
$\frac{d\dot{Q}}{dr}$ as a function of radius $r$ of star to show
the similar characters that the considerable modification of
Einstein's gravity and the global monopole with unified theory
order will raise the amount of $\frac{d\dot{Q}}{dr}$ greatly. The
stellar body with $f(R)$ global monopole can be well qualified as
a source of gamma-ray bursts. Moreover, we can select the factor
$\psi_{0}$ to be comparable with the accelerating universe while
regulate the parameter $\eta$ for the global monopole in order to
make the ratio curves to coincide with the results from astronomy.
It is possible to probe the monopole from astrophysical
observations.
\end{abstract}

\vspace{1cm} \hspace{0cm} PACS number(s): 04.70.Bw, 14.80.Hv\\
Keywords: Gamma-ray burst; $f(R)$ theory; global monopole

\noindent \textbf{I.\hspace{0.4cm}Introduction}

The temperature decreases during the evolution of universe [1, 2].
Various topological defects such as domain walls, cosmic strings
and monopoles may have been formed in the process of the vacuum
phase transition in the early stage of the universe [1, 2]. These
topological defects generated due to a breakdown of local or
global gauge symmetries. A global monopole is a spherically
symmetric topological defect arose in the phase transition of a
system composed of a self-coupling triplet of a scalar field whose
original global $O(3)$ symmetry is spontaneously broken to $U(1)$
[3]. There should exist a kind of massive sources involving global
monopoles and the metric of the source has a solid deficit angle
[4]. The spacetime metric with $f(R)$ modification is subject to
the fact of the accelerated expansion of the universe. The theory
of $f(R)$ gravity first proposed by Buchdahl [5] has been applied
to explain the accelerated-inflation problem instead of adding
dark energy or dark matter [6-8]. The $f(R)$ theory as a
generalization of the general relativity can be used to describe
the gravitational field of the massive object with a global
monopole [9]. It was discovered that the presence of the parameter
associated with the modification of gravity is indispensable in
providing stable circular orbits for particles [10]. The metric of
the gravitational object containing a global monopole within the
frame of $f(R)$ gravity has its own property [9-11]. The factor
from monopole for a typical grand unified theory changes its event
horizon [9, 10]. The non-vanishing modified parameter $\psi_{0}$
brings a cosmological horizon as a boundary of the universe to the
spacetime outside the source [9-11]. Certainly this kind of
metrics are composed of terms subject to the $f(R)$ model and
global monopoles [9-11]. Recently more investigations have been
paid for the black holes shown with this type of metric. We
studied the gravitational lensing of the Schwarzschild-like source
swallowing the $f(R)$ global monopole in the strong field limit
[12, 13]. We also calculated the thermodynamic quantities of this
kind of the black hole to examine the black hole's stability [14].
We derive the greybody factor for scalar fields in the
Schwarzschild spacetime with $f(R)$ global monopole [15]. The
timelike naked singularities of the black hole were discussed
[16]. The Hawking radiations of the $f(R)$ global monopole
corrected black hole was considered based on the Heisenberg
uncertainty principle or generalized uncertainty principle
respectively [17, 18]. Further, the absorption and scattering of
an $f(R)$-global-monopole black hole have been investigated [19].
It was found that the generalization of Heisenberg's uncertainty
principle leads the fragmentation of the $f(R)$-corrected black
hole involving global monopole [20]. We generalize the
consideration on the Hawking radiations and fragmentation of the
black hole to the case under the extended uncertainty principle to
show that the generalized uncertainty encourages the black hole
instability refer to the Parikh-Kraus-Wilczeck tunneling radiation
and division [21]. The observational appearances illuminated by
three simple models of accretions of an $f(R)$ global monopole
black hole were studied [22].

A lot of attentions from the astrophysical community have been
paid to an understanding of an energy source great enough to
explain the gamma-ray burst phenomenon. The hot accretion disk
emits neutrinos and antineutrinos [23-36]. Further the
neutrino-antineutrino annihilation into electrons and positrons
can become an energy source of gamma-ray bursts, so the gamma-ray
bursts may be thought as systems powered by newborn, stellar-mass
black holes accreting matter at hyper-critical rates [23-36]. The
processes $\nu+\bar{\nu}\longrightarrow e^{-}+e^{+}$ augment the
neutrino heating of the envelope leading a supernava explosion,
further the pair $e^{-}e^{+}$ near the surface of collapsing
neutron star sets free the gamma ray that may provide a possible
explanation of the observed bursts [37]. It was found that the
efficiency of neutrino-antineutrino annihilation into
electron-positron pair is enhanced over the Newtonian values up to
a factor of more than 4 in the case of Type II supernovae and by
up to a factor of 30 near the surface of collapsing neutron stars
[37]. The relativistic effects on the energy deposition rate
according to the neutrino pair annihilation near the rotation axis
of a Kerr black hole with a thin accretion disk was discussed [38,
39]. As the extension of works of Ref. [38, 39], the off-axis
contributions to the energy-momentum deposition rate from the
$\nu-\bar{\nu}$ pair collisions above a Kerr black hole within
thin accretion disk were probed [40]. The results of Ref. [40]
indicated that the off-axis energy deposition rate is larger by a
factor of 10-20 than the values in the on-axis cases. The authors
of Ref. [41] researched on the deposition of energy and momentum
by neutrino-antineutrino ($\nu\bar{\nu}$) annihilation in the
vicinity of an accretion disk or torus around a central
stellar-mass black hole to exhibit that the general relativity
effects and the rotation of Kerr spacetime both increase the
energy deposition rate by $\nu\bar{\nu}$-annihilation. The
neutrino-antineutrino annihilation into electron-positron pairs
near the surface of a neutron star within the frame of modified
gravity theories has been analyzed [42]. It was shown that the
energy deposition processes will increase significantly in the
neutron star and supernova envelope for charged Galileon, Einstein
dilation Gauss-Bonnet, Brans-Dicke, Eddington-inspired
Born-Infeld, Born-Infeld generalization of Reissner-Nordstrom
solution and higher derivative gravity, as various models of
gravity beyond the general relativity [42]. The authors of Ref.
[43] studied the influence of the presence of the quintessence
field around a gravitational object belonging to the black hole on
the neutrino pair annihilation efficiency like
$\nu+\bar{\nu}\longrightarrow e^{-}+e^{=}$ to show that the
quintessence powers the emitted energy rate ratio, so the
enhancement could be a source for the gamma-ray burst. Here we are
going to consider the neutrino pairs annihilation into
electron-positron pair ($\nu\bar{\nu}\longrightarrow e^{-}e^{+}$)
near the surface of neutron star or supernova including a global
monopole governed by $f(R)$ theory.

In this paper, we plan to discuss the energy deposition rate by
the neutrino annihilation process around the massive source
containing a global monopole in the context of $f(R)$ gravity. We
derive the integral form of the neutrino pair annihilation
efficiency based on the metric considered here. Secondly, we
calculate the ratio of total energy deposition to total Newtonian
energy deposition for factors $8\pi G\eta^{2}$ and $\psi_{0}$ from
global monopole and the correction to the general relativity
respectively. Our numerical estimation will reveal the influences
from global monopole and $f(R)$ approach on the possibility that
the astrophysical bodies attract the annihilation process
generating the gamma-ray burst. The results are listed in the end.

\vspace{0.8cm} \noindent \textbf{II.\hspace{0.4cm}The energy
deposition rate by the neutrino annihilation process in the
spacetime of massive source with an $f(R)$ global monopole}

We adopt the spherically symmetric line element as follow,

\begin{eqnarray}
\mathrm{ds}^{2}=g_{\mu\nu}dx^{\mu}dx^{\nu}\hspace{4cm}\nonumber\\
=A(r)dt^{2}-B(r)dr^{2}-r^{2}(d\theta^{2}+\sin^{2}\theta
d\varphi^{2})
\end{eqnarray}

In the $f(R)$ gravity theory, the action is given by [9-11],

\begin{equation}
S=\frac{1}{2\kappa}\int d^{4}x\sqrt{-g}f(R)+S_{m}
\end{equation}

\noindent where $f(R)$ is an analytical function of Ricci scalar
$R$, $\kappa=8\pi G$ with $G$ Newton constant, and $g$ the
determinant of the metric tensor (1). Here $S_{m}$ is the action
associated with the matter fields as global monopole [5],

\begin{equation}
S_{m}=\int
d^{4}x\sqrt{-g}[\frac{1}{2}(\partial_{\mu}\phi^{a})(\partial^{\mu}\phi^{a})
-\frac{1}{4}\lambda(\phi^{a}\phi^{a}-\eta^{2})^{2}]
\end{equation}

\noindent where the triplet of the field configuration showing a
monopole is [2, 3],

\begin{equation}
\phi^{a}=\eta h(r)\frac{x^{a}}{r}
\end{equation}

\noindent with $x^{a}x^{a}=r^{2}$. Here, $\lambda$ and $\eta$ are
model parameters. This model has a global $O(3)$ symmetry, which
is spontaneously broken to $U(1)$ [2, 3].

With help of the Einstein Equation, the field equation reads [5,
44-46],

\begin{equation}
F(R)R_{\mu\nu}-\frac{1}{2}f(R)g_{\mu\nu}-\nabla_{\mu}\nabla_{\nu}F(R)
+g_{\mu\nu}F(R)=\kappa T_{\mu\nu}
\end{equation}

\noindent where $F(R)=\frac{df(R)}{dR}$, and $T_{\mu\nu}$ is the
minimally coupled energy-momentum tensor from action (3). Under
the weak field approximation, the components of metric tensor can
be chosen as $A(r)=1+a(r)$ and $B(r)=1+b(r)$ with $|a(r)|$ and
$|b(r)|$ being smaller than unity [9-11]. The field equation is
solved and the metric is found [9-11],

\begin{eqnarray}
A(r)=B^{-1}(r)\hspace{2cm}\nonumber\\
=1-8\pi G\eta^{2}-\frac{2GM}{r}-\psi_{0}r
\end{eqnarray}

\noindent Here, the generalized theory of gravity corresponds to a
tiny correction to the general relativity like $F(R(r))=1+\psi(r)$
with $\psi(r)<<1$ [9-11]. The deviation of standard general
relativity can be taken as the simplest analytical function of the
radial coordinate $\psi(r)=\psi_{0}r$ [9-11]. It should be pointed
that the correction $\psi_{0}r$ in the metric (6) is linear, which
is different from those in cases such as de Sitter spacetime and
the Reissner-Nordstrom metric, etc.. In addition, the monopole
parameter $\eta$ is of the order $10^{16}GeV$ according to the
typical grand unified theory, which means $8\pi G\eta^{2}\approx
10^{-5}$ [3]. For metric (6), as roots of $A(r)=0$, a cosmological
horizon [9-11],

\begin{equation}
r_{c}=\frac{1}{\psi_{0}}(1-8\pi G\eta^{2}+\sqrt{(1-8\pi
G\eta^{2})^{2}-8GM\psi_{0}})
\end{equation}

\noindent appears beside an event horizon,

\begin{equation}
r_{h}=\frac{1}{\psi_{0}}(1-8\pi G\eta^{2}-\sqrt{(1-8\pi
G\eta^{2})^{2}-8GM\psi_{0}})
\end{equation}

\noindent The presence of a nonzero $\psi_{0}$ bring a
cosmological horizon as a boundary of the universe to the
spacetime described by the $f(R)$ monopole metric, but the
spacetime without gravity modification is asymptotically flat.

We plan to treat the energy deposition in the spacetime with the
descriptions of (1) and (6). The energy deposition per unit time
and per volume for the neutrino annihilation process is given by
[37, 47],

\begin{equation}
\frac{dE(\mathbf{r})}{dtdV}=2KG_{F}^{2}F(r)\int\int
n(\varepsilon_{\nu})n(\varepsilon_{\bar{\nu}})
(\varepsilon_{\nu}+\varepsilon_{\bar{\nu}})
\varepsilon_{\nu}^{3}\varepsilon_{\bar{\nu}}^{3}d\varepsilon_{\nu}
d\varepsilon_{\bar{\nu}}
\end{equation}

\noindent where

\begin{equation}
K=\frac{1}{6\pi}(1\pm 4\sin^{2}\theta_{w}+8\sin^{4}\theta_{w})
\end{equation}

\noindent with the Weinberg angle $\sin^{2}\theta_{w}=0.23$ and,

\begin{equation}
K(\nu_{\mu},\bar{\nu}_{\mu})=K(\nu_{\tau},\bar{\nu}_{\tau})
=\frac{1}{6\pi}(1-4\sin^{2}\theta_{w}+8\sin^{4}\theta_{w})
\end{equation}

\begin{equation}
K(\nu_{e},\bar{\nu}_{e})=\frac{1}{6\pi}(1+4\sin^{2}\theta_{w}+8\sin^{4}\theta_{w})
\end{equation}

\noindent and the Fermi constant $G_{F}=5.29\times
10^{-44}cm^{2}MeV^{-2}$. The angular integration factor is
represented by [37],

\begin{eqnarray}
F(r)=\int\int(1-\Omega_{\nu}\Omega_{\bar{\nu}})^{2}
d\Omega_{\nu}d\Omega_{\bar{\nu}}\nonumber\\
=\frac{2\pi^{2}}{3}(1-x)^{4}(x^{2}+4x+5)\hspace{1cm}
\end{eqnarray}

\noindent where

\begin{equation}
x=\sin\theta_{r}
\end{equation}

\noindent The angle $\theta_{r}$ is between the particle
trajectory and the tangent vector to a circular orbit at radius
$r$. $\Omega_{\nu}(\Omega_{\bar{\nu}})$ is the unit direction
vector and $d\Omega_{\nu}(d\Omega_{\bar{\nu}})$ is a solid angle.
Here $n(\varepsilon_{\nu})$ and $n(\varepsilon_{\bar{\nu}})$ are
number densities in phase space [37],

\begin{equation}
n(\varepsilon_{\nu})=\frac{2}{h^{3}}\frac{1}
{e^{\frac{\varepsilon_{\nu}}{kT}}+1}
\end{equation}

\noindent where $h$ is Planck constant. The integration of Eq.(9)
can be performed and the expression of rate per unit time and unit
volume is given by [37],

\begin{equation}
\frac{dE}{dtdV}=\frac{21\zeta(5)\pi^{4}}{h^{6}}KG_{F}^{2}
F(r)(kT)^{9}
\end{equation}

The local temperature measured by a local observer is defined as
[37, 47],

\begin{equation}
T(\mathbf{r})\sqrt{g_{00}(\mathbf{r})}=constant
\end{equation}

\noindent The neutrino temperature at the neutrinosphere reads
[37],

\begin{equation}
T(r)\sqrt{g_{00}(r)}=T(R)\sqrt{g_{00}(R)}
\end{equation}

\noindent where $g_{00}$ is a component of spacetime metric. The
luminosity relating to the redshift can be selected as [37],

\begin{equation}
L_{\infty}=g_{00}(R)L(R)
\end{equation}

\noindent where the luminosity for a single neutrino species at
the neutrinosphere is [37],

\begin{eqnarray}
L(R)=L_{\nu}+L_{\bar{\nu}}\nonumber\\
=4\pi R^{2}\frac{7}{4}\frac{ac}{4}T^{4}(R)
\end{eqnarray}

\noindent where $a$ is the radiation constant and $c$ is the speed
of light in vacuum. The Eq.(18), Eq.(19) and Eq.(20) are
substituted into the Eq.(16) to obtain [37],

\begin{equation}
\frac{dE(\mathbf{r})}{dtdV}=\frac{21\zeta(5)\pi^{4}}{h^{6}}
KG_{F}^{2}k^{9}(\frac{7}{4}\pi
ac)^{-\frac{9}{4}}L_{\infty}^{\frac{9}{4}}F(r)
\frac{(g_{00}(R))^{\frac{9}{4}}}{(g_{00}(r))^{\frac{9}{2}}}R^{-\frac{9}{2}}
\end{equation}

\noindent In order to calculate the angular integration $F(r)$, we
should further the discussion on the variable $x$. We follow the
procedure of Ref.[37],

\begin{equation}
\tan\theta=\sqrt{\frac{|g_{11}|}{|g_{33}|}}\frac{dr}{d\varphi}
\end{equation}

The null geodesic in the spacetime of a spherically symmetric
gravitational object is given by [44-47],

\begin{equation}
(\frac{1}{r^{2}}\frac{dr}{d\varphi})^{2}+\frac{A(r)}{r^{2}}
=\frac{1}{b^{2}}
\end{equation}

\noindent where $b$ is the impact parameter. The expression (22)
can be substituted into Eq.(23) and simplified to give [37],

\begin{equation}
\frac{r^{2}}{A(r)}\cos^{2}\theta=b^{2}
\end{equation}

\noindent The components $g_{11}$ and $g_{33}$ are from metric (1)
and (6). According to Eq.(24), it can be rewritten [42, 48],

\begin{eqnarray}
\frac{r^{2}}{A(r)}\cos^{2}\theta_{r}
=\frac{R^{2}}{A(R)}\cos^{2}\theta_{R}=b^{2}\nonumber\\
\cos^{2}\theta_{r}=\frac{R^{2}}{r^{2}}\frac{A(r)}{A(R)}
\cos^{2}\theta_{R}\hspace{1cm}
\end{eqnarray}

\noindent Combining Eq.(14) and Eq.(25), it can be inserted [42,
48],

\begin{eqnarray}
x^{2}=\sin^{2}\theta_{r}|_{\theta_{R}=0}\nonumber\\
=1-\frac{R^{2}}{r^{2}}\frac{A(r)}{A(R)}\hspace{0.5cm}
\end{eqnarray}

\noindent We can proceed the integration of rate per unit time and
unit volume shown in Eq.(21) [43],

\begin{eqnarray}
\dot{Q}=\frac{dE}{\sqrt{g_{00}}dt}\hspace{9.5cm}\nonumber\\
=4\pi\int_{R}^{R_{CH}}r^{2}\sqrt{-g_{11}}(\frac{dE}{dtdV})dr
\hspace{6.6cm}\nonumber\\
=\frac{84\zeta(5)\pi^{5}}{h^{6}}KG_{F}^{2}k^{9}(\frac{7}{4}\pi
ac)^{-\frac{9}{4}}L_{\infty}^{\frac{9}{4}}(g_{00}(R))^{\frac{9}{4}}
R^{-\frac{9}{2}}\int_{R}^{R_{CH}}\frac{r^{2}\sqrt{-g_{11}}
F(r)}{(g_{00}(r))^{\frac{9}{2}}}dr
\end{eqnarray}

\noindent where $R_{CH}$ is the cosmological horizon. Here
$\dot{Q}$ can reflect the total amount of energy converted from
neutrinos to electron-positron pairs at any radius [37]. According
to the energy deposition rate (27) and the quantities from
Ref.[37, 42, 43], it is possible to write [42],

\begin{equation}
\frac{\dot{Q}}{\dot{Q}_{Newt}}=3(g_{00}(R))^{\frac{9}{4}}
\int_{1}^{\frac{R_{CH}}{R}}(x-1)^{4}(x^{2}+4x+5)
\frac{y^{2}\sqrt{-g_{11}(Ry)}}{(g_{00}(Ry))^{\frac{9}{2}}}dy
\end{equation}

\noindent with dimensionless variable $y=\frac{r}{R}$.

We can also obtain the function of radius $r$ like
$\frac{d\dot{Q}}{dr}$ to reflect the enhancement according to the
issue of Ref.[37],

\begin{eqnarray}
\frac{d\dot{Q}}{dr}=4\pi(\frac{dE}{dtdV})\sqrt{-g_{11}}r^{2}
\hspace{5cm}\nonumber\\
=\frac{168\zeta(5)\pi^{7}}{3h^{6}}KG_{F}^{2}k^{9}(\frac{7}{4}\pi
ac)^{-\frac{9}{4}}L_{\infty}^{\frac{9}{4}}\hspace{3.5cm}\nonumber\\
\times(x-1)^{4}(x^{2}+4x+5)
(\frac{g_{00}(R)}{g_{00}(r)})^{\frac{9}{4}}R^{-\frac{5}{2}}
\sqrt{-g_{11}}(\frac{r}{R})^{2}
\end{eqnarray}

It is significant to discuss the ratio (28) in the background
described by metric components (6). Comparing the metric (1) with
the component function (6), we can relate that $g_{00}(r)=A(r)$
and $g_{11}(r)=-B(r)=-\frac{1}{A(r)}$ to obtain,

\begin{equation}
g_{00}(r)=-\frac{1}{g_{11}(r)}=1-8\pi G\eta^{2}-\frac{2GM}{r}
-\psi_{0}r
\end{equation}

\noindent We rewrite the variable (26) around the gravitational
source with solid deficit angle in the context of $f(R)$ theory,

\begin{equation}
x^{2}=1-\frac{1}{y^{2}}\frac{1-8\pi
G\eta^{2}-\frac{2GM}{R}\frac{1}{y}-\psi_{0}Ry} {1-8\pi
G\eta^{2}-\frac{2GM}{R}-\psi_{0}R}
\end{equation}

\noindent We substitute the metric (30) into the Eq.(28) to obtain
the energy deposition rate for the neutrino annihilation
surrounding the source involving the $f(R)$ global monopole,

\begin{equation}
\frac{\dot{Q}}{\dot{Q}_{Newt}}=3(g_{00}(R))^{\frac{9}{4}}
\int_{1}^{\frac{r_{H}}{R}}(x-1)^{4}(x^{2}+4x+5)
\frac{y^{2}}{(g_{00}(Ry))^{5}}dy
\end{equation}

\noindent It should be emphasized that,

\begin{equation}
g_{00}(R)=1-8\pi G\eta^{2}-\frac{2GM}{R}-\psi_{0}R
\end{equation}

\noindent and

\begin{equation}
g_{00}(Ry)=1-8\pi G\eta^{2}-\frac{2GM}{R}\frac{1}{y}-\psi_{0}Ry
\end{equation}

It is necessary to quantify the integral expression of ratio
$\frac{\dot{Q}}{\dot{Q}_{Newt}}$ by Eq.(32) and depict the
dependence on $\frac{R}{M}$ in the Figures. We compare the curves
under the conditions to find that more considerable deviations
from the general relativity give rise to larger ratio
$\frac{\dot{Q}}{\dot{Q}_{Newt}}$ shown in Figure 1. Meanwhile it
is found that the larger values of the parameter like $\psi_{0}$
result in much more amount of the ratio, which is similar to the
conclusions from Ref.[42, 43]. It is obvious that the existence of
global monopoles also augment the quotient of $\dot{Q}$ and
$\dot{Q}_{Newt}$ according to the Figure 2. It should be pointed
out that the larger values of parameters describing the monopoles
cause more ratio. As an interesting example, we plot the emitted
energy ratio $\frac{\dot{Q}}{\dot{Q}_{Newt}}$ which is relevant
for the generation of gamma-ray burst due to the monopole
parameter $8\pi G\eta^{2}\approx 10^{-5}$ from typical grand
unified theory [3] and the generalization of general relativity
$\psi_{0}\approx 10^{-3}$ [9-11, 22] in the Figure 3. It is
indicated that the converting energy per unit time for the
annihilation is made larger by a factor almost up to 2 at about
$R\approx10M$. In the case of Type II supernova with nearly
$R\approx5M$, the factor grows to be more than 5, which is enough
to explain the observed GRBs. Within the range down to
$R\approx3M$ corresponding to collapsing neutron stars, the factor
for the promotion becomes to be more than 10. According to Ref.
[49], We can also explore this kind of massive source supporting
the burst in another direction. At first, the value of $\psi_{0}$
can be chosen under the limit of the observing data for
accelerating universe. Adjusting the global monopole variable
$\eta$ to enable the curves of $\frac{\dot{Q}}{\dot{Q}_{Newt}}$ to
coincide the ones from observation, we can estimate $\eta$ by
comparing with the order from unified theory [3]. It is clear that
the annihilations surrounding the celestial body swallowing the
$f(R)$ global monopole will emit much greater energy per unit
time, which is the distinct feature of this kind of gamma-ray
burst.

For the sake of stressing the promotion of $e^{-}e^{+}$ pair
energy from the neutrino annihilation, we start to elaborate the
$\frac{d\dot{Q}}{dr}$ as a function of radius for several stellar
masses $\frac{M}{R}$. We illustrate the dependence of the
derivative on the dimensionless variable $\frac{r}{R}$ because of
$8\pi G\eta^{2}\approx 10^{-5}$ and $\psi_{0}\approx 10^{-3}$ as
mentioned above [9-11, 22] in Figure 4. It is similar that the
structure with smaller values of $\frac{R}{M}$ result in the
greater $\frac{d\dot{Q}}{dr}$. Certainly the augment is much
stronger near the surface of the neutron star like Ref.[37]. It
should be emphasized that the annihilation of
neutrino-antineutrino pairs owing to the Schwarzschild-like
spacetime modified by the presence of global monopole in view of
$f(R)$ theory may provide the much more energy than the needed to
efficiently originate the gamma-ray bursts. It is not necessary
for the burst to occur near the surface of the specific source
from Figure 4.

\vspace{0.8cm} \noindent \textbf{III.\hspace{0.4cm}Conclusion}

Our investigations are performed in this paper refer to the
neutrino pair annihilation $\nu\bar{\nu}\longrightarrow
e^{-}e^{+}$ around the gravitational source involving the global
monopole in the frame of $f(R)$ approach. We calculate the emitted
energy rate ratio $\frac{\dot{Q}}{\dot{Q}_{Newt}}$ in the allowed
range of parameters for $f(R)$ gravity and global monopole in the
background of a massive source with an $f(R)$ global monopole to
find that the $f(R)$ corrections to the general relativity can
also stimulate the annihilation process to produce excessive
energy per unit time, while exhibit that the existence of global
monopole can significantly increase much more energy deposition.
The neutrino pair annihilation subject to this kind of
gravitational bodies emits much greater energy per unit time to be
proposed as a well-qualified candidate of gamma-ray bursts no
matter the central objects are Type II supernova or collapsing
neutron stars. By comparison of our calculations with the
observational results, we can open a new window to estimate the
global monopole parameter.

\vspace{1cm}
\noindent \textbf{Acknowledge}

This work is partly supported by the Shanghai Key Laboratory of
Astrophysics.

\newpage
\begin{figure}
\setlength{\belowcaptionskip}{10pt} \centering
\includegraphics[width=15cm]{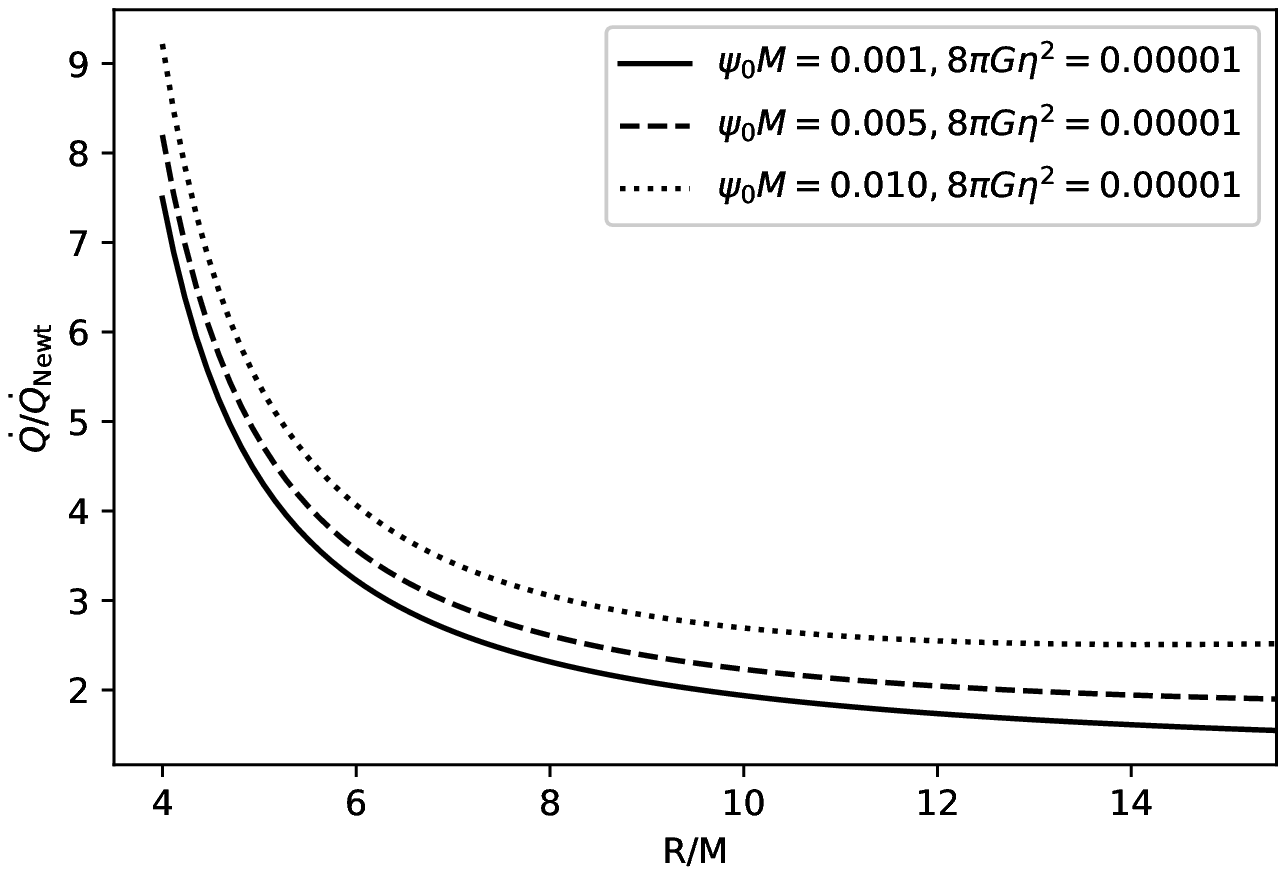}
\caption{The solid, dotted and dashed curves of the ratio
$\frac{\dot{Q}}{\dot{Q_{Newt}}}$ as functions of radius of
gravitational sources for $f(R)$ factors $\psi_{0}=0.001, 0.005,
0.01$ respectively with global monopole variable $8\pi
G\eta^{2}=10^{-5}$}
\end{figure}

\newpage
\begin{figure}
\setlength{\belowcaptionskip}{10pt} \centering
\includegraphics[width=15cm]{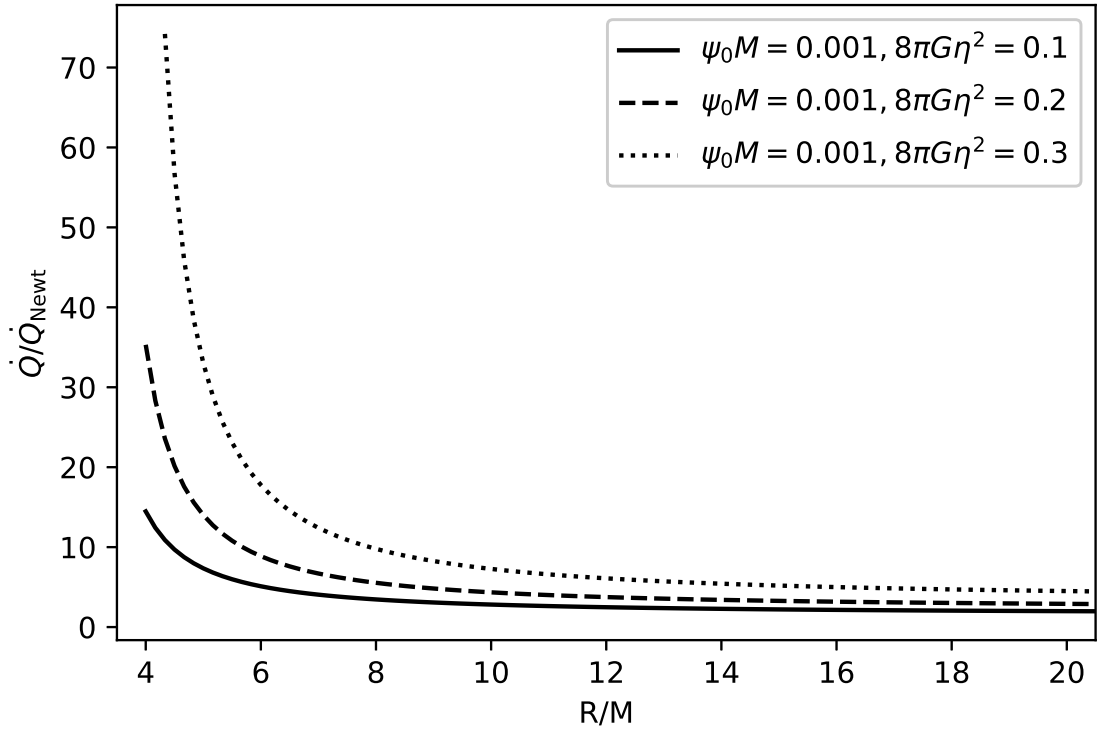}
\caption{The solid, dotted and dashed curves of the ratio
$\frac{\dot{Q}}{\dot{Q_{Newt}}}$ as functions of radius of
gravitational sources for global monopole variable $8\pi
G\eta^{2}=0.1, 0.2, 0.3$ respectively with $f(R)$ factors
$\psi_{0}=0.001$}
\end{figure}

\newpage
\begin{figure}
\setlength{\belowcaptionskip}{10pt} \centering
\includegraphics[width=15cm]{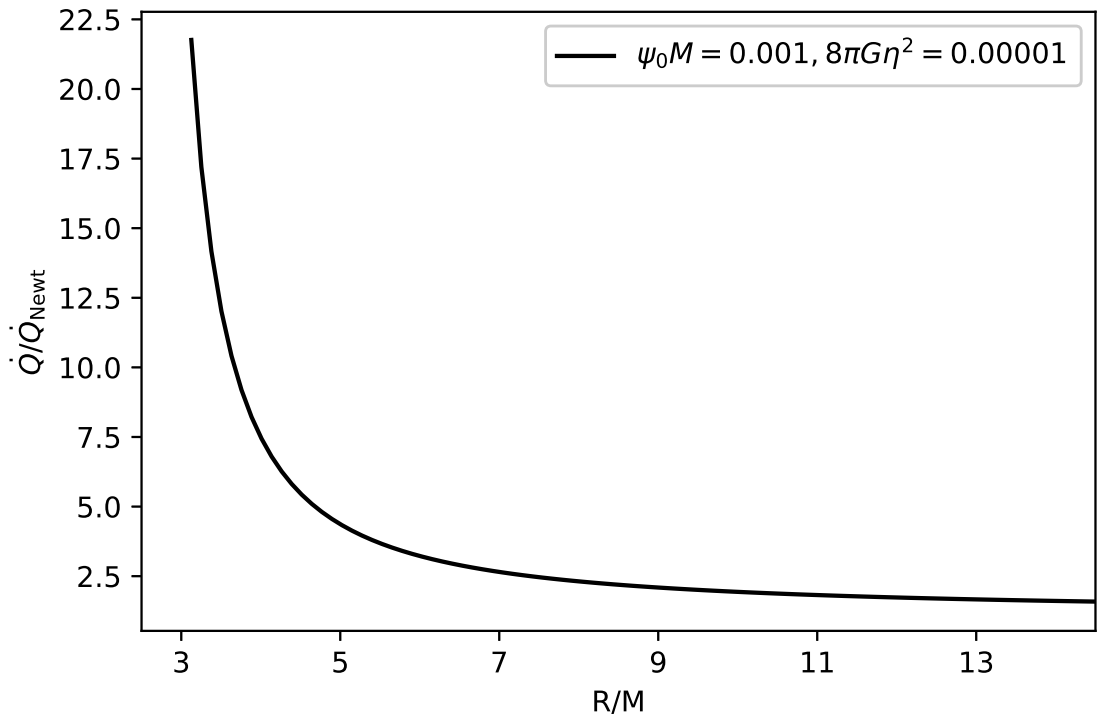}
\caption{The curve of the ratio $\frac{\dot{Q}}{\dot{Q_{Newt}}}$
as functions of radius of gravitational source for $f(R)$ factors
$\psi_{0}=0.001$ and global monopole variable $8\pi
G\eta^{2}=10^{-5}$ $f(R)$}
\end{figure}

\newpage
\begin{figure}
\setlength{\belowcaptionskip}{10pt} \centering
\includegraphics[width=15cm]{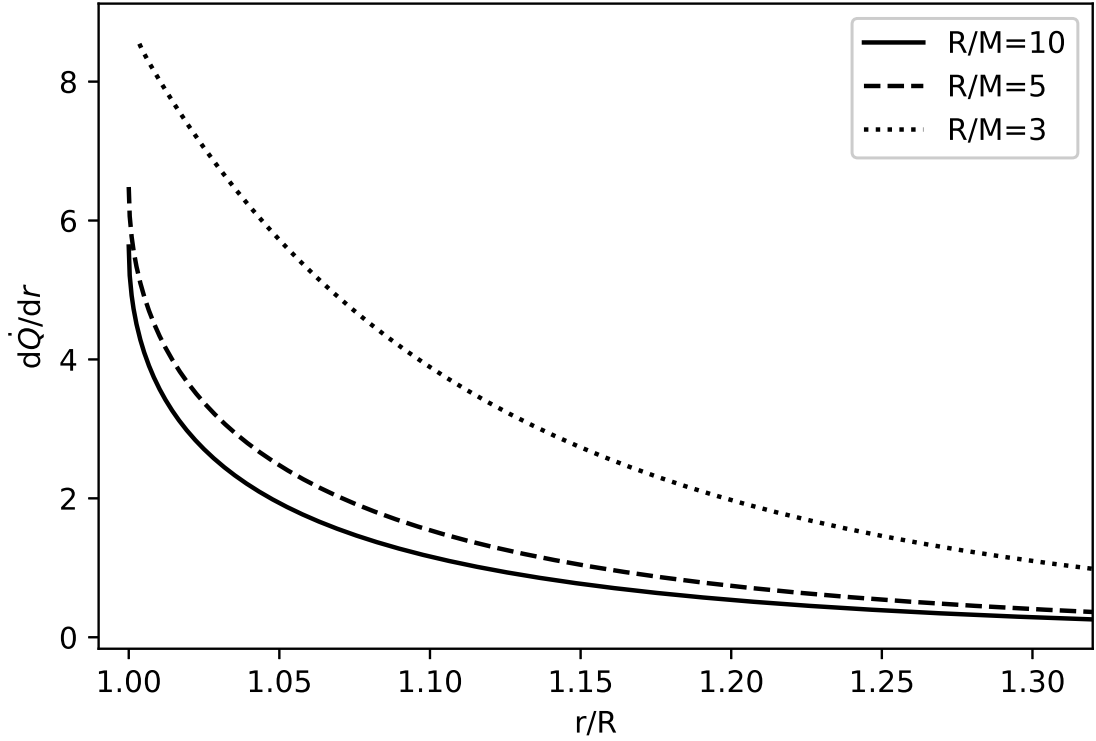}
\caption{The dotted, dashed and solid curves of the derivative
$\frac{d\dot{Q}}{dr}$ as function of radius of various
gravitational source for $f(R)$ factor $\psi_{0}=0.001$ and global
monopole variable $8\pi G\eta^{2}=10^{-5}$ under $\frac{R}{M}=3,
5, 10$ respectively}
\end{figure}

\end{document}